\documentclass[10pt,twocolumn,a4paper]{IEEEtran}

\usepackage{amsmath}
\usepackage[utf8]{inputenc}
\usepackage[english]{babel}
\usepackage{siunitx}
\usepackage{hyperref}
\usepackage{authblk}
\usepackage{lipsum,multicol}

\usepackage{romannum}

\hypersetup{colorlinks=true, pdfstartview=FitV, linkcolor=red, citecolor=blue, plainpages=false, urlcolor=blue}

\usepackage{amssymb}
\usepackage{graphicx}

\usepackage[algo2e,ruled,vlined]{algorithm2e}
\SetKwInput{KwInput}{Input}                
\SetKwInput{KwOutput}{Output}              

\usepackage{float}
\usepackage{titlesec}
\titleformat{\subsection}[runin]
       {\normalfont\bfseries}
       {\thesubsection}
       {0.5em}
       {}
       [.]

\title{Large field-of-view non-invasive imaging through scattering layers using fluctuating random illumination}

\author[1,2]{Lei Zhu} 
\author[1]{Fernando Soldevila}
\author[1]{Claudio Moretti}
\author[1]{Alexandra d'Arco}
\author[1]{Antoine Boniface}
\author[2]{Xiaopeng Shao}
\author[1]{Hilton B. de Aguiar}
\author[1,*]{Sylvain Gigan}

\affil[1]{Laboratoire Kastler Brossel, ENS--Université PSL, CNRS, Sorbonne Université, College de France, 24 Rue Lhomond, F-75005 Paris, France.}
\affil[2]{School of Physics and Optoelectronic Engineering, Xidian University, Xi'an 710071, China}
\affil[*]{Corresponding author: sylvain.gigan@lkb.ens.fr}

\begin{document}
\twocolumn[
\begin{@twocolumnfalse}
\maketitle

\begin{abstract}
Non-invasive optical imaging techniques are essential diagnostic tools in many fields. Although various recent methods have been proposed to utilize and control light in multiple scattering media, non-invasive optical imaging through and inside scattering layers across a large field of view remains elusive due to the physical limits set by the optical memory effect, especially without wavefront shaping techniques. Here, we demonstrate an approach that enables non-invasive fluorescence imaging behind scattering layers with field-of-views extending well beyond the optical memory effect. The method consists in demixing the speckle patterns emitted by a fluorescent object under variable unknown random illumination, using matrix factorization and a novel fingerprint-based reconstruction. Experimental validation shows the efficiency and robustness of the method with various fluorescent samples, covering a field of view up to three times the optical memory effect range. Our non-invasive imaging technique is simple, neither requires a spatial light modulator nor a guide star, and can be generalized to a wide range of incoherent contrast mechanisms and illumination schemes. 
\end{abstract}
\end{@twocolumnfalse}
]

\section*{\label{sec:level0} INTRODUCTION}
Non-invasive optical imaging has important applications in various fields ranging from biotechnology \cite{zhao_non-invasive_2001,artzi_vivo_2011} to optical detection \cite{kozloff_non-invasive_2009}. However, inhomogeneous samples, such as biological tissues, scatter light, which results in a complex speckle pattern on the detector \cite{Goodman1976,Bender}. With increasing depth, separating the low amount of ballistic light from the scattered light becomes a big challenge \cite{Abramson1978,huang_optical_1991}. Over the years, many approaches have been put forward to overcome this problem by exploiting or suppressing the scattered light. With the development of spatial light modulators (SLMs), multiple ways to control and manipulate scattered light have been developed \cite{Mosk2012,rotter_light_2017}. Several techniques have been proposed to focus light by making use of feedback signals to optimize the incident wavefront to recreate a focus that is then used for raster-scanning microscopy \cite{Vellekoop2007,Horstmeyer2015}. These techniques require access to both sides of the scattering layer to optimize the wavefront, which strongly limits their application in real-case scenarios. To overcome this, other strategies have been proposed based on wavefront shaping and various feedback signals such as fluorescence or ultrasound signals \cite{Horstmeyer2015,Katz2019,Popoff2010,Hofer2019}. However, these approaches either require long acquisition times or are limited to small fields of view (FoV). On the other hand, several techniques exploiting the angular speckle correlations, known as the optical memory effect (ME) \cite{Freund1988,Yllmaz2019,Osnabrugge}, have also been proposed for imaging objects hidden behind scattering media \cite{Bertolotti2012,Katz2014}. While these approaches are fast, their FoV is still limited by the ME range. 

Linear fluorescence is widely used in biology and biomedical sciences \cite{Ruan2020,Lichtman2005,mangeat_super-resolved_2021}. It enables imaging of cellular, subcellular, or molecular components, and has the advantages of high spatial resolution, contrast, and speed. Recent advances have allowed both focusing and imaging through scattering media using fluorescent light. Even so, these methods either rely on the use of guide stars \cite{Horstmeyer2015}, are limited to the ME range \cite{Hofer2018}, or need to characterize the scattering medium \cite{Boniface2020}.

Here, we present a robust approach that allows to non-invasively image through scattering layers at depths far beyond the ME range, using fluctuating random speckle illumination simply generated with a rotating diffuser. Once excited, each fluorescent emitter generates a unique speckle pattern on the detector, which constitutes its fingerprint. Each image captured by the camera is an incoherent sum of the fingerprints from the emitters, with different relative weights due to the variable random illumination patterns and sample structure. To retrieve each individual fingerprint, we capture a set of images while randomly changing the illumination, and use a Non-negative Matrix Factorization (NMF) algorithm to demix the set of acquired frames. After that, the fingerprints are used to reconstruct the final image by exploring the correlations between them. To validate the technique, we experimentally demonstrate our non-invasive approach both on fluorescent beads and on continuous fluorescent objects.

\section*{\label{sec:level1} RESULTS}

\begin{figure*}[ht!]
\includegraphics[width=\textwidth]{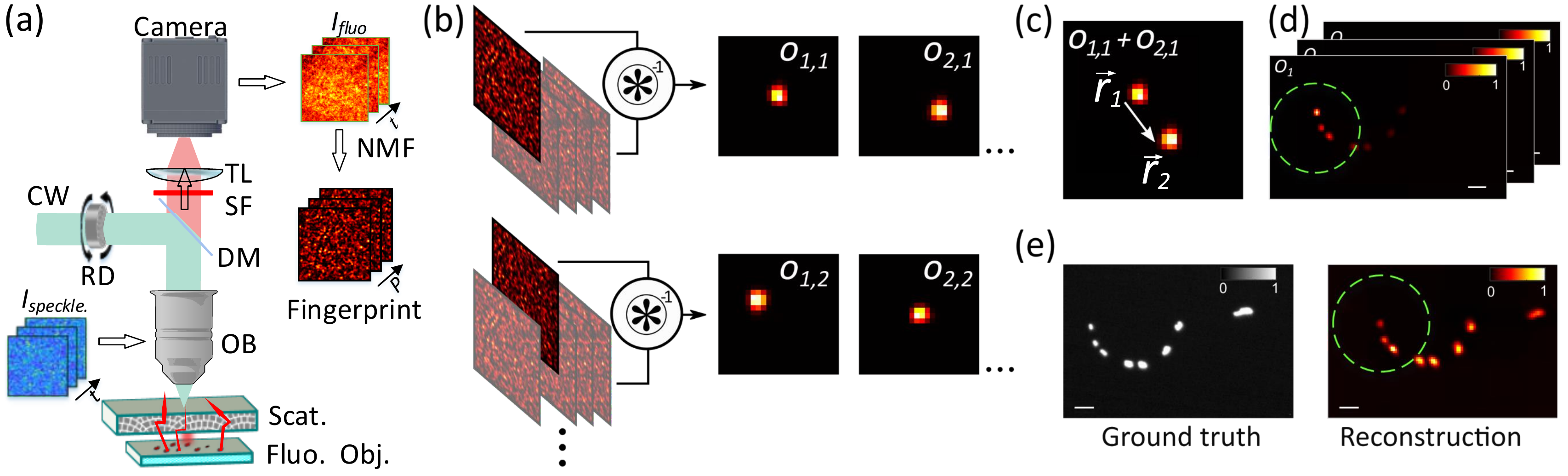}
\caption{\label{fig:1}Schematic of the experimental setup and reconstruction principle. (a) Schematic view of experimental setup. A coherent light source illuminates a rotating diffuser in order to excite the fluorescent object through a scattering medium with a random modulated speckle pattern. Once excited, the emitted signal from the fluorescent objects is recorded with a camera. $I_{fluo}$ is a series of $t$ fluorescent speckles corresponding to different random speckle illuminations. The fingerprints can be recovered from $I_{fluo}$ by using NMF. (b-e) Fingerprint-based reconstruction. (b) Pairwise deconvolution (labelled as $\circledast^{-1}$) between all the possible pairs of emitter fingerprints is performed. (c) The result of each deconvolution provides the relative position between one emitter and its neighbors. (d) By adding the resulting images for each emitter, it is possible to recover a partial image of the object centered at that emitter (see Eq.\ref{eq:4}). (e) All the partial images can be merged into the final reconstruction according to the relative position between neighbouring emitters. Dashed circle indicates the optical memory range. Scale bar sizes are $\SI{10}{\micro\meter}$. RD: rotating diffuser, DM: dichroic mirror, OB: objective, Scat.: scattering medium, Fluo. Obj.: fluorescent object, SF: spectral filter, TL: tube lens.}
\end{figure*}

The experimental setup is depicted in Fig.\ref{fig:1}a. A rotating holographic diffuser modulates the incident light coming from a laser by adding a random phase when the light propagates through it. Then, the modulated light travels through the scattering medium and generates a random unknown speckle which illuminates the object. The excitation speckle induces a fluorescent response from the object, which propagates back through the scattering medium, generating an incoherent sum of speckles that is measured by the camera. Although the captured images are low-contrast, random, and seemingly information-less, they contain all the fingerprints from the independent emitters of the object, but with time-varying weights. Furthermore, independent emitters within the ME range will produce correlated but shifted fingerprints on the camera \cite{Freund1988}, while emitters outside the ME range will produce totally uncorrelated fingerprints. For a given speckle illumination, the captured image, $I_{\textsl{fluo}}$, can be expressed as a linear superposition of those fingerprints with different weights. Thus, the camera image is given by:

\begin{equation}
I_{fluo}(r,t) = \sum^{P}_{k=1} w_{k}(r) h_{k}(t),
\label{eq:1}
\end{equation}
where $I_{fluo}(r,t)$ corresponds to a low contrast speckle for the $t^{th}$ illumination, $r$ is the spatial coordinate, $w_{k}(r)$ represents the fingerprint of the $k^{th}$ independent emitter of the object, $h_{k}(t)$ stands for the amount of excitation light at the $k^{th}$ emitter during the $t^{th}$ illumination, and $P$ is the number of independent emitters. Given enough different random illuminations, a collection of frames can be used to retrieve each individual fingerprint, $w_{k}(r)$, by using an NMF algorithm, that we will now explain in detail.

\subsection*{\label{sec:level21}Fingerprint demixing procedure}

After randomly exciting the object with a variety of $t$ random speckles, a series of camera images, $I_{fluo}(r,t)$, are collected. It is possible to retrieve the fingerprints, $w_{k}$, corresponding to each independent emitter, from the measurements $I_{fluo}(r,t)$ by finding the solution to the minimization problem \cite{Berry2007}:

\begin{equation}
\min_{W>0,H>0} \Arrowvert I -WH\Arrowvert^{2}
\label{eq:2},
\end{equation}

This minimization problem can be formulated as a low rank factorization, where the matrix $I \in \mathbb{R}_{+}^{r \times t} $ contains all the $I_{fluo}(r,t)$, can be approximated with two real positive matrices $W \in \mathbb{R}_{+}^{r \times \rho} $ (the fingerprints) and $H \in \mathbb{R}_{+}^{\rho \times t}$ (the temporal evolutions), where $r$ are the pixels, $\rho$ is the estimated rank of $I$ and $t$ indicates the frames. Since the collected images and the demixed fingerprints are positive, this problem corresponds exactly to the family of NMF problems. The NMF framework has been employed in demixing scenarios, both in structural imaging \cite{ Boniface2020} and functional imaging \cite{Moretti2020a,Pegard2016, Pnevmatikakis2016}. In our case, the estimated rank $\rho$ approximately corresponds to the number of independent emitters $P$ and it can be estimated from the data by minimizing the root mean squared residual of NMF as a function of the rank (see Supplementary \Romannum{1}).

\subsection*{\label{sec:level22} Fingerprint-based reconstruction}

After the demixing step, the fingerprints are retrieved. Due to the ME, emitters close to each other will produce highly correlated fingerprints, with a spatial shift that is directly determined by their relative position \cite{Katz2014}. By exploring these correlations, a position map of the emitters can be recovered, thus yielding an image of the object. Several approaches can be used to explore the correlations and calculate the shifts between fingerprints. Usually, this process is performed by doing a cross-correlation between the fingerprints and locating the position of the maximum \cite{Boniface2020}. However, here we introduce a novel approach based on deconvolution, that we denote as Fingerprint-based Reconstruction (FBR). Compared to the cross-correlation procedure, we found that this approach allows to suppress noise and strongly improves the quality of the reconstruction.

No matter if two emitters are within the same ME patch or not, one can perform the pairwise deconvolution of the $i$\textsl{-th} emitter by the $k$\textsl{-th} emitter, which can be written as:

\begin{equation}
\underset{o_{i,k}}{\arg\min}
\frac{\mu}{2} \vert\vert w_{i}-o_{i,k}\circledast w_{k}\vert\vert^2_{2}+\vert\vert o_{i,k}\vert\vert_{TV}
\label{eq:3},
\end{equation}
where $\mu$ is a regularization parameter, $\circledast$ denotes the convolution operator, $\vert\vert \cdot \vert\vert_{2}$ indicates the $L_{2}$ norm and $\vert\vert \cdot \vert\vert_{TV}$ represents the Total Variation (TV) norm, and the two fingerprints are $w_{i}$ (considered as the "image")  and $w_{k}$ (considered as the "point spread function" (PSF)). When the two emitters lay within one ME range, the pairwise deconvolution yields a uniform image with a narrow delta-like peak, which is located at a distance from the center given by the relative position of the two emitters ($\vec{r}_{i,k} = \vec{r}_i - \vec{r}_k$). If the two emitters are located beyond the ME range, the deconvolution yields noise.

For a given emitter $k$, it is possible to obtain, $O_{k}$, the partial image of the object in the vicinity of the emitter, by simply adding the result of all the pairwise deconvolutions related to that emitter, $o_{i,k}$. (see Fig.\ref{fig:1}b-d).

\begin{equation}
O_{k} = \sum^{\rho}_{i=1} 
o_{i,k}
\label{eq:4},
\end{equation}

Even if the ensemble of emitters expands well beyond the ME range, the full spatial distribution can be recovered if the different isoplanatic patches are "connected" by emitters (see Fig.\ref{fig:1}d). For example, if emitters $i$ and $k$ are beyond the ME range but emitter $j$ is between them, we can always calculate the shift between them as $\vec{r}_{i,k} = \vec{r}_{i,j} + \vec{r}_{j,k}$. The global reconstruction $O^{Global}$ can be obtained by composing all the partial images, $O_{k}$, into one image, taking into account their relative positions with respect to the first emitter, $\vec{r}_{k,1}$: 
\begin{equation}
O^{Global} = \sum^{\rho}_{k=1} 
O_{k}(\vec{r} - \vec{r}_{k,1})
\label{eq:5},
\end{equation} 

\begin{figure}[ht]
\includegraphics[width=\linewidth]{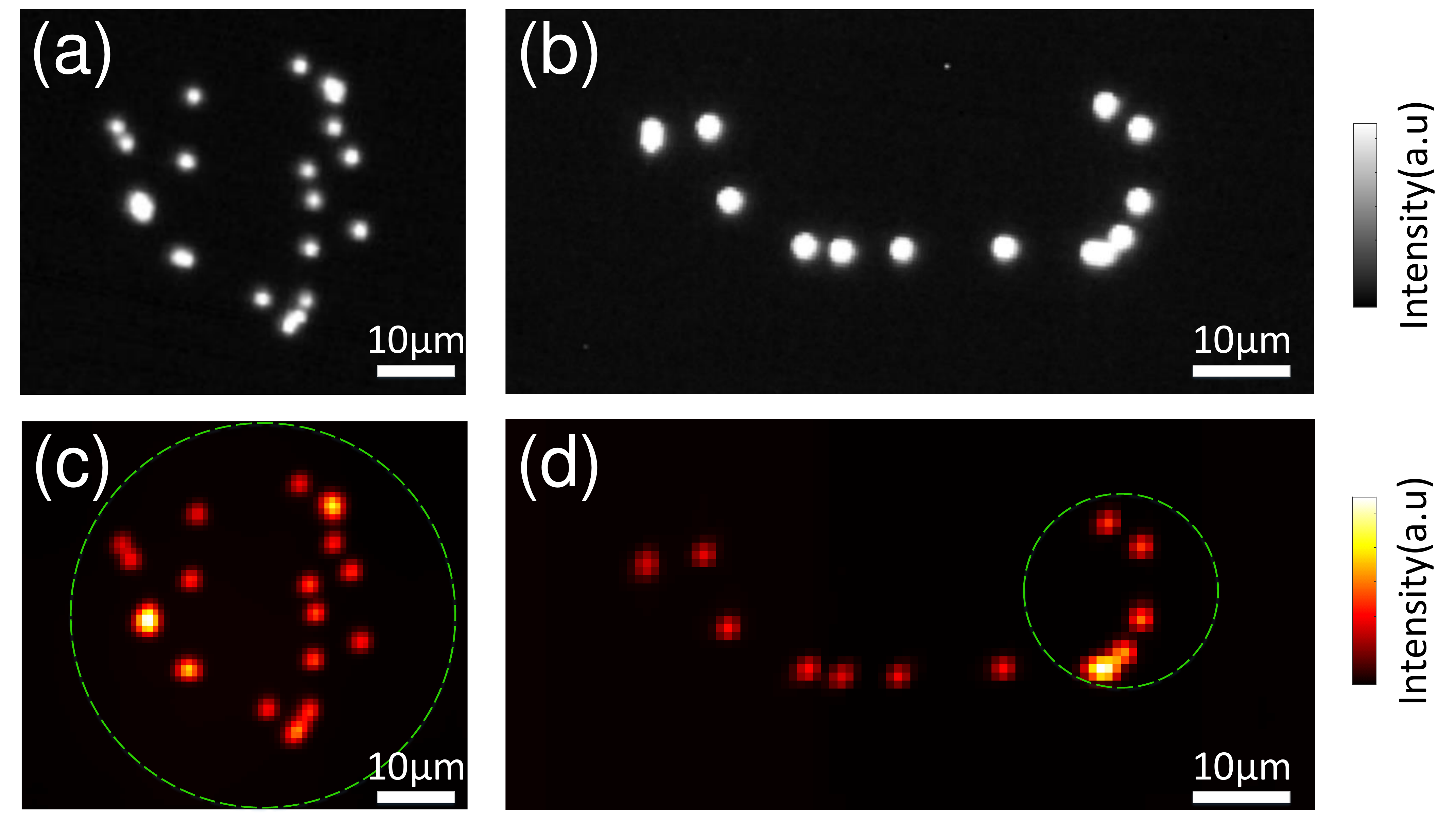}
\caption{\label{fig:2} Experimental results of imaging through a scattering medium with fluorescent beads. (a)-(b) Fluorescent images of beads recorded without scattering medium. (c)-(d) Reconstruction of the object using NMF+FBR approach. The estimated rank of NMF is $\rho = 26$ for (c) and $\rho = 16$ for (d). In both cases, $t = 5120$ fluorescent speckle patterns are captured. The exposure time of (c) and (d) is set to $\SI{15}{\milli\second}$ and $\SI{20}{\milli\second}$, respectively. Dashed circles indicate the optical memory effect range.}
\end{figure}

\begin{figure}[ht]
\includegraphics[width=\linewidth]{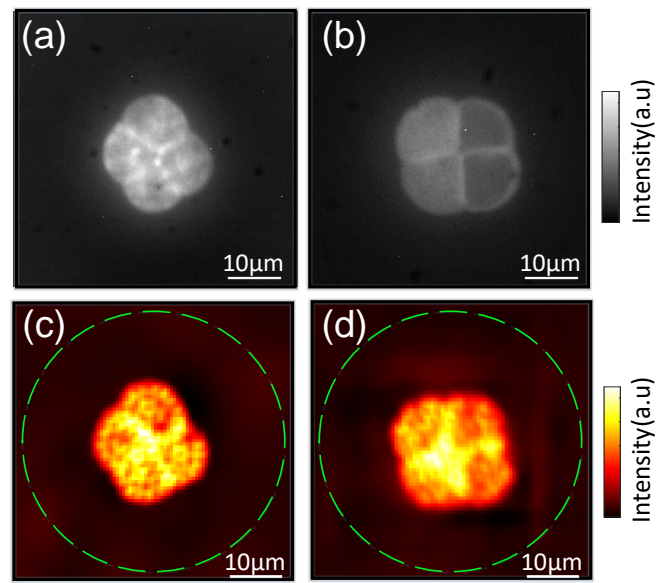}
\caption{\label{fig:3} Experimental results of imaging through scattering media with continuous objects. (a)-(b) Fluorescent images of different pollen seed structures recorded without scattering medium. (c)-(d) Reconstruction of the object with the NMF+FBR approach. The estimated rank for the NMF is $\rho = 68$ for (c) and $\rho = 85$ for (d), respectively. In both cases, $t = 5120$ fluorescent speckle patterns are recorded with an exposure time of $\SI{10}{\milli\second}$. Dashed circle indicates the optical memory effect range.}
\end{figure}

Experimentally, we prove that our technique can be used to recover very sparse objects by using 2D distributions of beads with a diameter of $\SI{1}{\micro\meter}$. As shown in Fig.\ref{fig:2}, our approach can reconstruct the objects without constraints from the ME range. Non-sparse, continuous objects are common in scattering biological samples, and this often poses a difficult challenge for non-invasive imaging through scattering media approaches. To demonstrate that our technique can also work with non-sparse and continuous objects, we use fluorescence-stained pollen grains, whose reconstructed images are shown in Fig.\ref{fig:3}. The size of fluorescence-stained pollen grains in Fig.\ref{fig:3} is smaller than the ME range and the process of reconstruction used for them is the same as what is presented in Fig.\ref{fig:1}.

In both Fig.\ref{fig:2} and Fig.\ref{fig:3} we report the reconstruction of hidden beads distributions and pollen seeds. However, the more complex the objects are, the larger the required number of independent illuminations. Indeed, given more speckle patterns, NMF is able to provide more reliable fingerprints, thus a more reliable reconstruction (See supplementary Fig.\ref{fig:10}). We note that a few hundred illuminations are sufficient to recover the object reasonably well in our case. Importantly, our technique is not limited by the number of independent illuminations that we can generate with the rotating diffuser, as it is possible to produce a very large number of independent illumination with different rotating diffusers by tunning their scattering angle \cite{Pnevmatikakis2016}. As an alternative to a rotating diffuser, we also propose a version of the setup using a SLM, which allows reproducible pattern projections without practical limitations on the number of patterns (see Supplementary \Romannum{4}).

\section*{Discussion}

Here, we report on successfully recovering the hidden objects by exploiting the correlation between the fingerprints. However, we believe, based on the spectral ME or the 3D ME, that the technique could be used to recover multi spectral or 3D objects. This is a significant advantage as it does not require the scattering medium to be static. Another important future direction will be to explore the ME of dynamic scattering media to recover the hidden object through it. Interestingly, our technique, contrarily to the techniques based on the optical transmission matrix, does not require focusing light on the object, and relies only upon the random video frames generated from the random illumination. Another important point is the simplicity of the technique, which does not require calibration of the PSFs \cite{Antipa2018} of the imaging system, and can be implemented without the need of expensive SLMs. Thus, our technique can be easily implemented in various scattering media and imaging scenarios. It is necessary to stress that our approach reconstructs both non-sparse and continuous objects over previous autocorrelation approaches \cite{Bertolotti2012,Katz2014}.  

In conclusion, we have shown a non-invasive technique to computationally retrieve images of objects hidden behind a scattering medium from low-contrast fluorescent speckles using random illumination. We have demonstrated that our approach works with both sparse objects, even beyond the ME range, and continuous objects. Importantly, the proposed approach neither relies upon ballistic light nor uses wavefront shaping, and it is adaptable to various scattering media and objects. Our technique is flexible, robust, and opens a promising avenue towards deep fluorescence imaging in highly scattering media. Finally, it can be generalized to a wide range of incoherent contrast mechanisms and illumination schemes.

\section*{METHODS}
\subsection*{Experimental setup}

A continuous-wave laser ($\lambda  = \SI{532}{\nano\meter} $, Coherent Sapphire) is expanded and illuminates the rotating holographic diffuser (Edmund, DG10). Then the modulated light is delivered onto the fluorescent sample through a $\SI{200}{\milli\meter}$ lens (LA1708-A, Thorlabs) and objective (Zeiss W “Plan-Apochromat” $\times20$, NA $1.0$). After excitation, the fluorescence is scattered by the medium and collected with a $\SI{150}{\milli\meter}$ tube lens (L, AC254-150-A, Thorlabs), which is employed to produce an image onto the detector, a sCMOS camera (Hamamatsu ORCA Flash). Two dichroic filters (short pass $\SI{532}{\nano\meter} $, Thorlabs  and $\SI{533}{\nano\meter} $ notch MF525-39, Thorlabs) are used to block any signal that does not come from the fluorescence emission. The fluorescent objects, which are made of orange beads ($ \SI{540/560}{\nano\meter} $, Invitrogen FluoSpheres, size $\SI{1.0}{\micro\meter}$) or pollen seeds (Carolina, Mixed Pollen Grains Slide, $w.m.$), are placed below the scattering medium. The distance between the scattering medium and the fluorescent objects is $\SI{0.2}{\milli\meter}$. A transmission pathway that consists of a microscope objective (Olympus ''MPlan N'' $\times20$, NA $0.4$), a $\SI{150}{\milli\meter}$ tube lens (L, AC$254-150-$A, Thorlabs), and CCD camera (Allied Vision, Manta), is used as a passive control only. This control part is used to correctly select the position of fluorescent object with a white light source (Moritex, MHAB 150W) and it also allows us to align the experimental setup. For the scattering medium, we either use a single holographic diffuser (Newport, 10DKIT-C1,$\SI{10}{\degree}$) or a combination of two (Newport, 10DKIT-C1,$\SI{10}{\degree}$ and 10DKIT-C1,$\SI{1}{\degree}$) in order to get different memory effect ranges.

The exposure time has been set from $\SI{10}{\milli\second}$ to $\SI{20}{\milli\second}$, depending on the scattering medium and the fluorescent sample. Once captured, the speckle images, which contain few tens of speckle grains, are cropped from the raw images. Then, a high-pass Fourier Gaussian filter is employed to remove the background from the cropped images and the processed data set is analyzed with the NMF algorithm to obtain the fingerprint of each emitter. The experimental setup is shown in Supplementary \Romannum{5}.

In Fig.\ref{fig:2}a-b, the size of each cropped image is $70\times72$ pixels, the number of patterns $t$ is 5120, the scattering medium is a holographic diffuser (Newport, 10DKIT-C1, $10$), and the exposure time is $\SI{15}{\milli\second}$. In Fig.\ref{fig:2}c-d, the size of each cropped image is $64\times66$ pixels, the number of patterns $t$ is 5120, the scattering medium are two holographic diffusers (Newport, 10DKIT-C1, $10$ + $1$), and the exposure time is $\SI{20}{\milli\second}$. In Fig.\ref{fig:3}, the size of each cropped image is $70\times74$ pixels, the number of patterns $t$ is 5120, the holographic diffuser (Newport, 10DKIT-C1, $10$) is used as the scattering medium, and the exposure time is $\SI{10}{\milli\second}$.

\subsection*{NMF+fingerprint-based reconstruction algorithm}

For the NMF, knowing the rank of the system is necessary. The rank $\rho$ is estimated by looking at the root mean square residual $\vert\vert I_{fluo}-WH \vert\vert_{Fro}$ as a function of rank $\rho$ (detailed in Supplementary \Romannum{1}), where $\vert\vert \cdot \vert\vert_{Fro}$ indicates the Frobenius norm, and minimizing it. For the NMF, a random initialization is employed. The retrieved fingerprints are used as the input data of the FBR. 




\subsection*{Acknowledgements}
This research has been funded by the FET-Open (Dynamic-863203), European Research Council ERC Consolidator Grant (SMARTIES-724473), China Scholarship Council (CSC) (201906960055). 

\clearpage
\section*{Supplementary information}

\section{Estimating the rank of the system}\label{Supplementary A}

\begin{figure}[h]
\includegraphics[width=\linewidth]{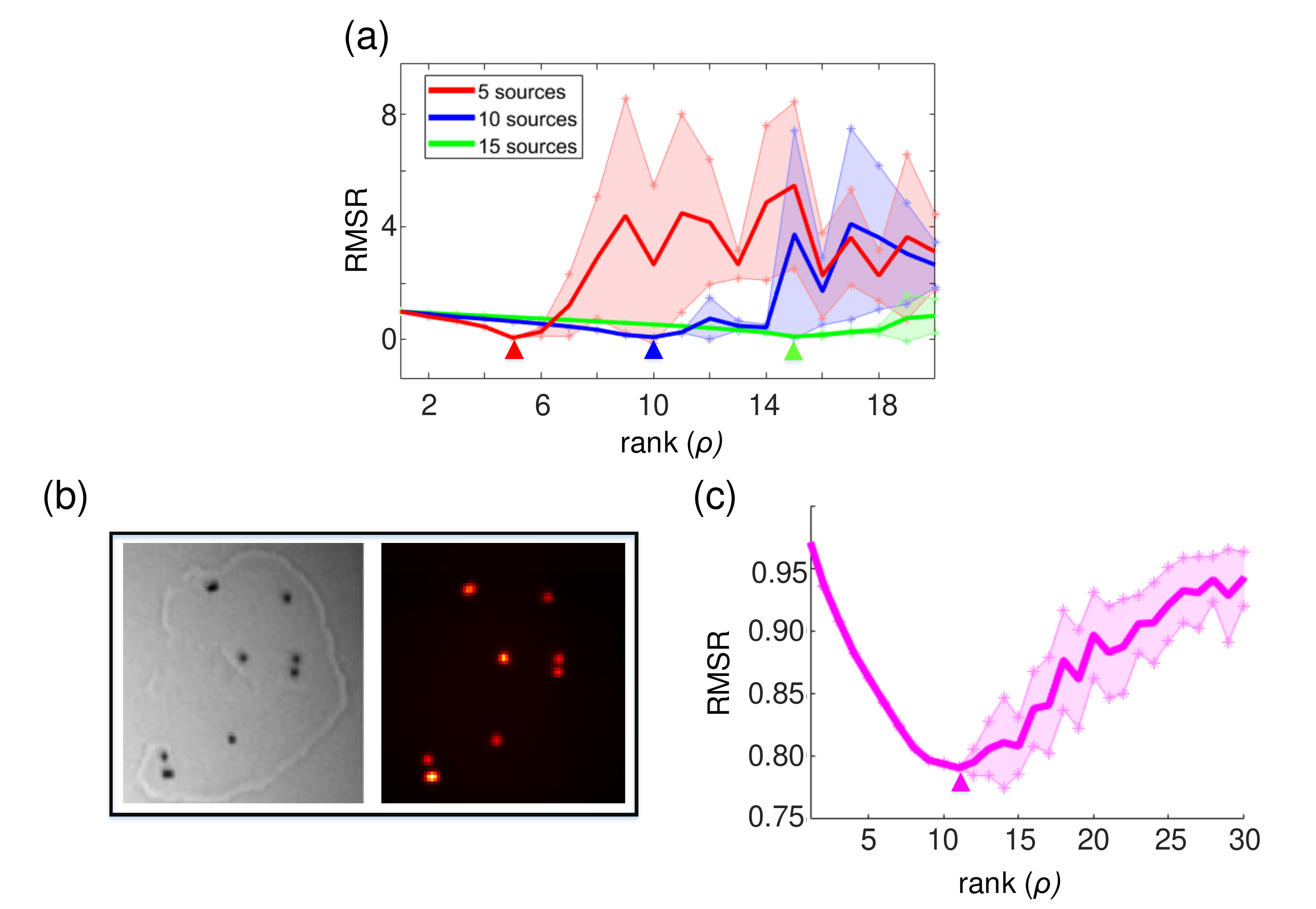}
\caption{\label{fig:4} NMF rank estimation procedure. (a) Simulation results with $P$ fluorescent sources. The triangle symbol marks the minimum mean value of the RMSR, the solid line stands for the mean value of the RMSR over 12 times with the random initialization value of NMF, and the asterisk is the error bar that indicates its standard deviation in both positive and negative direction. The minimum mean RMSR value is achieved when the estimated rank $\rho$ equals the true rank $P$ of the system. (b) Experimental results with a fluorescent object which contains $P = 11$ beads. (c) Estimating the rank of a fluorescent object shown in (b), the solid line stands for the mean value of the RMSR over 12 repetitions, the triangle symbol marks the minimum mean value of the RMSR, and the error bars indicate the standard deviation of the RMSR.}
\end{figure}

In order to operate the NMF algorithm, the rank, $\rho$, is a unique parameter which needs to be pre-determined. Several methods can be used to estimate this rank, such as looking at clusters using a \textsl{k}-means algorithm \cite{Moretti2020a}, or by estimating the 'elbow' of the loss function \cite{Boniface2020,hutchins_position-dependent_2008}. In our case, we use a simple procedure based on the study of the evolution of the root mean square residual (RMSR) of the NMF for different ranks to estimate the true rank of the system. Once we acquire a dataset, it is possible to demix it by setting any desired rank $\rho$. For each $\rho$, we can estimate the quality of the NMF solution by the RMSR: $\vert\vert I_{fluo}-WH \vert\vert_{Fro}$. As we change the rank, the RMSR varies, reaching a minimum when the estimated rank equals the true rank of the system. Also, when overestimating the rank, it is possible to have a big variance on the result of the NMF for multiple realizations. Both these behaviors can be used to delimit the true value of the rank of the system. As an example of the procedure, we show both a simulation with different numbers of beads and the same procedure with experimental datasets in Fig.\ref{fig:4}. As can be seen in Fig.\ref{fig:4}a, the minimum value of RMSR is $\rho = 5,10$, and $15$, respectively, which is in excellent agreement with the ground truth number of sources ($5,10$, and $15$). The rank estimation procedure is performed 12 times with different random initializations in order to see the variance of the results.

\begin{figure}[h]
\includegraphics[width=\linewidth]{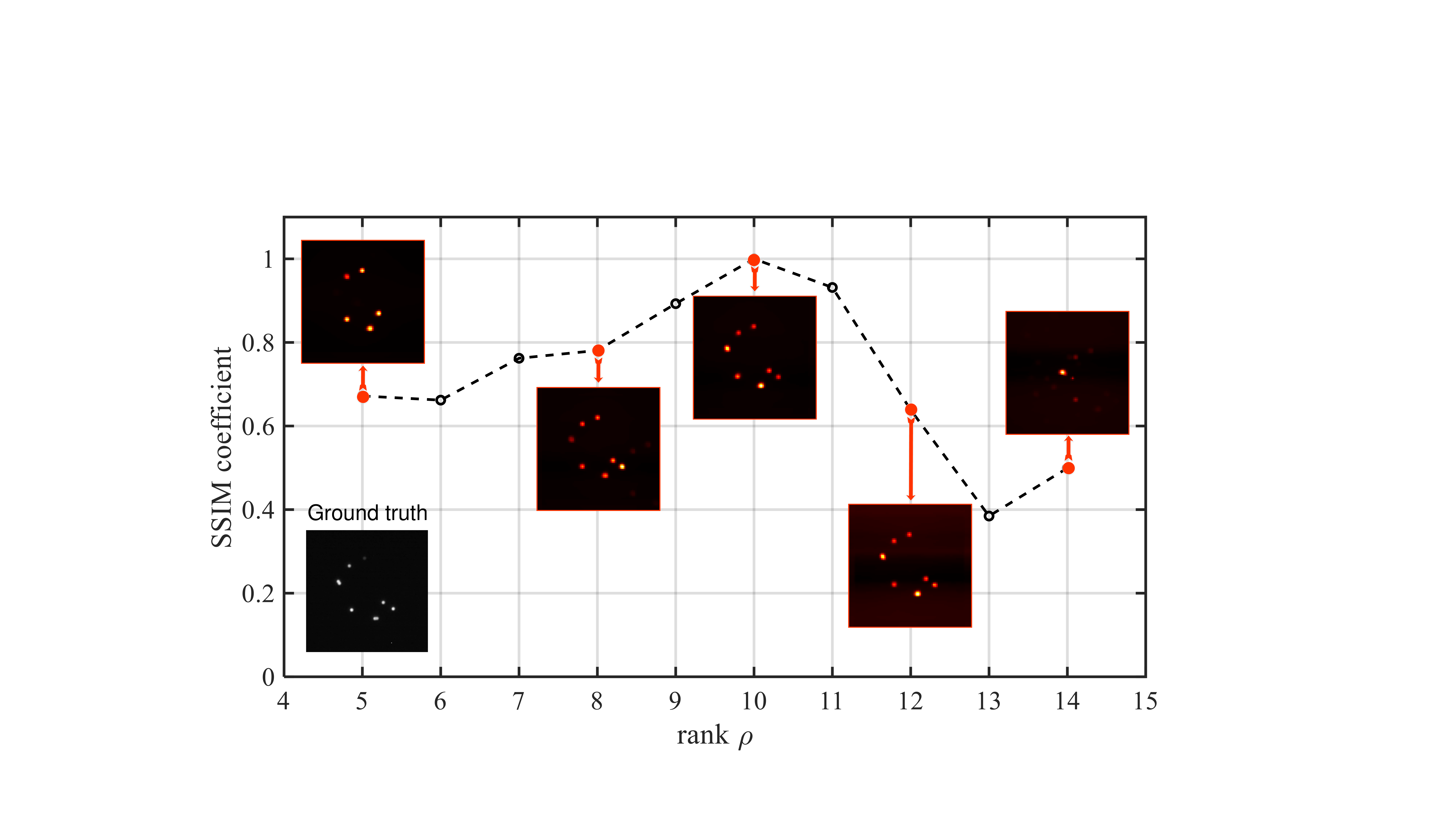}
\caption{\label{fig:5} The plot of SSIM coefficient between different reconstructions and reference image. The reconstruction corresponding to $\rho = 10$ is treated as a reference image (the number of beads in the sample is $10$).}
\end{figure}

Furthermore, we investigate the effect of incorrectly estimating the ranks $\rho$ of the system on the quality of our reconstruction. We experimentally choose a fluorescent object which contains 10 beads. Then, we recover the object using our approach with different rank estimations, ranging from 6 to 15. The structural similarity index metric (SSIM) \cite{Daoud2017} is used to quantify the performance of reconstruction. As shown in Fig.\ref{fig:5}, our technique is capable of achieving reconstruction with different ranks even if the proposed rank is not fully accurate. However, underestimating the rank may lead to losing some beads in the global reconstruction. On the other hand, overestimating the rank on the NMF will produce mixed patterns that hinder the capability to obtain the relative position between different emitters.  

\section{Optical memory effect}\label{Supplementary B}

\begin{figure}[h]
\includegraphics[width=\linewidth]{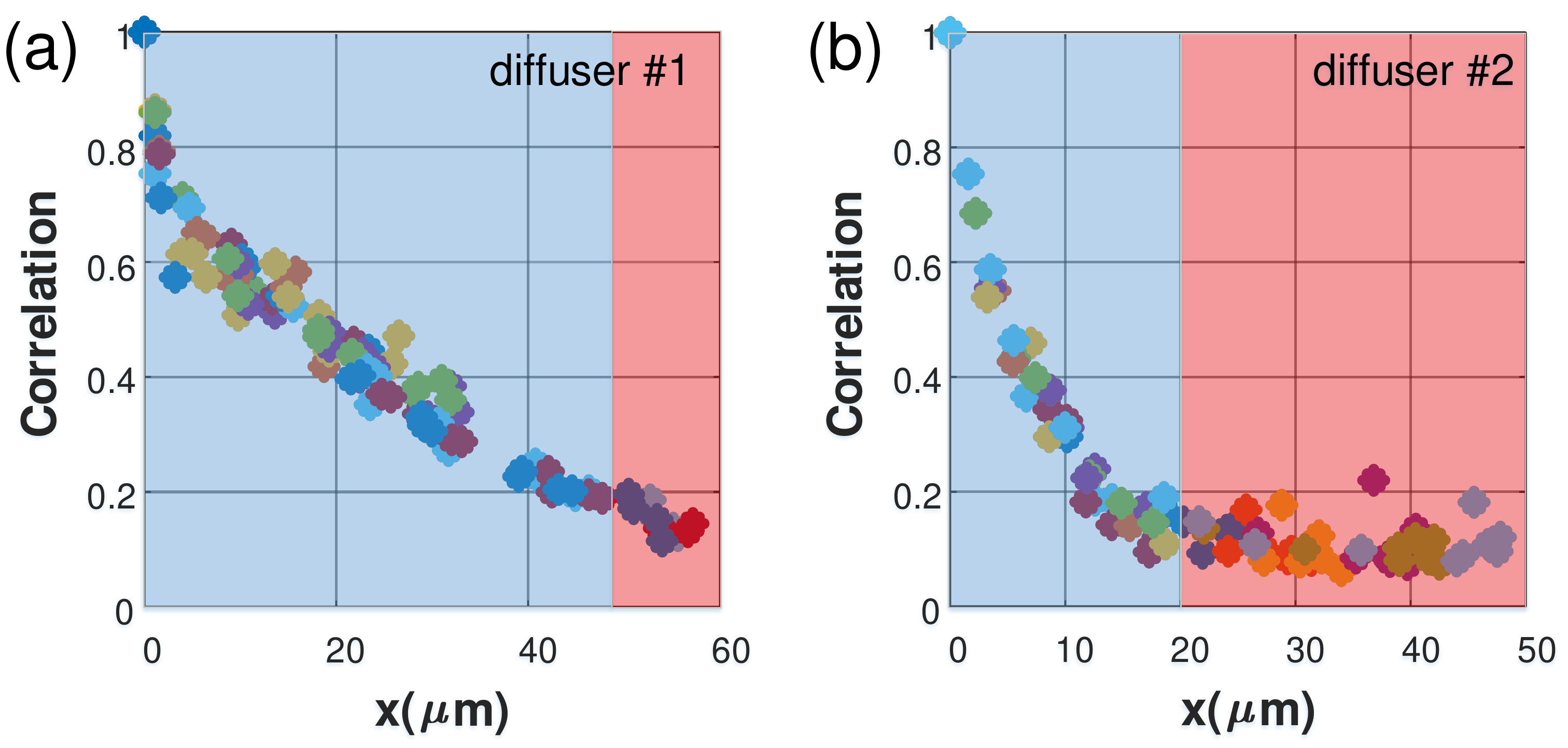}
\caption{\label{fig:6} Plot of maximum value of cross-correlations between different patterns corresponding to different fluorescent bead positions for diffuser $\#1$ (a) and $\#2$ (b), where the colorcode indicates the different data sets.}
\end{figure}

With the purpose of proving that our technique can be applied in imaging beyond the ME, different diffusers with different scattering properties are tested. We estimate the ME range of each diffuser by exploring the correlation between the speckle patterns generated while laterally moving an emitter (a small fluorescent bead). As the source moves, the correlation between speckle patterns gradually decreases, reaching a minimum correlation when the distance is longer than the ME range. Measuring this distance allows estimating the ME range for each diffuser. As reported in Fig.\ref{fig:6}, the ME range is approximately $\SI{50}{\micro\meter}$ and $\SI{20}{\micro\meter}$ corresponding to diffuser $\#1$ and diffuser $\#2$, respectively. Our non-invasive technique not only can retrieve objects within ME range as shown in Fig.\ref{fig:2}a of the manuscript, which is possible to be recovered via autocorrelation approaches, but also can reconstruct an extended object, which could not be retrieved with autocorrelation approaches. As shown in Fig.\ref{fig:2}b of the manuscript, the size of the reconstructed extended object is around $\SI{58}{\micro\meter} \times \SI{22}{\micro\meter}$ with the diffuser $\#2$.

\section{Fingerprint-based reconstruction}\label{Supplementary c}

\begin{figure}[h!]
\includegraphics[width=\linewidth]{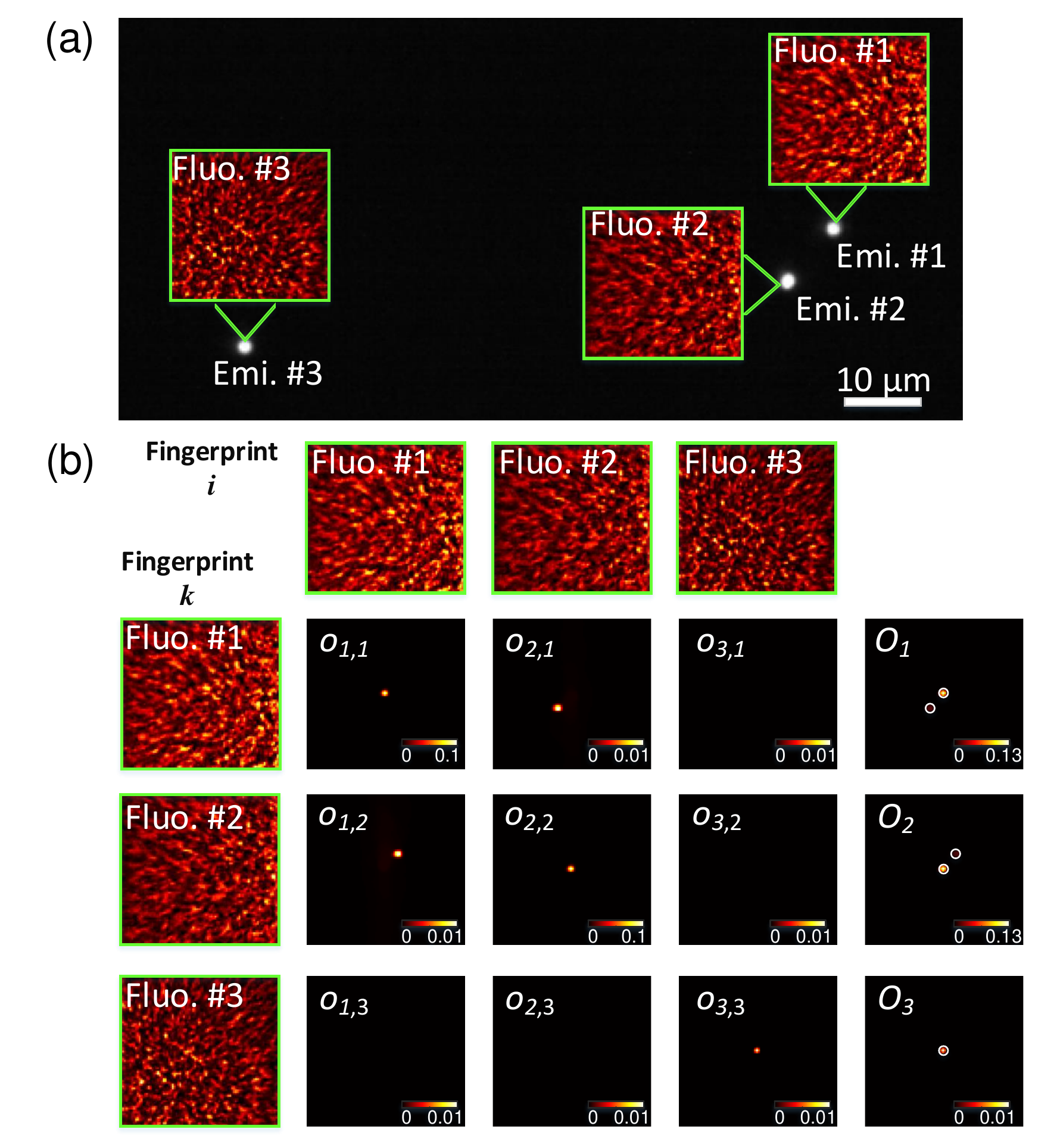}
\caption{\label{fig:7} Detail of FBR. (a) Ground truth taken without diffuser. (b) The detail of pairwise deconvolution. The estimated rank $\rho$ of this data set is 3 and the exposure time is $\SI{10}{\milli\second}$.}
\end{figure}

\begin{figure}[h!]
\includegraphics[width=\linewidth]{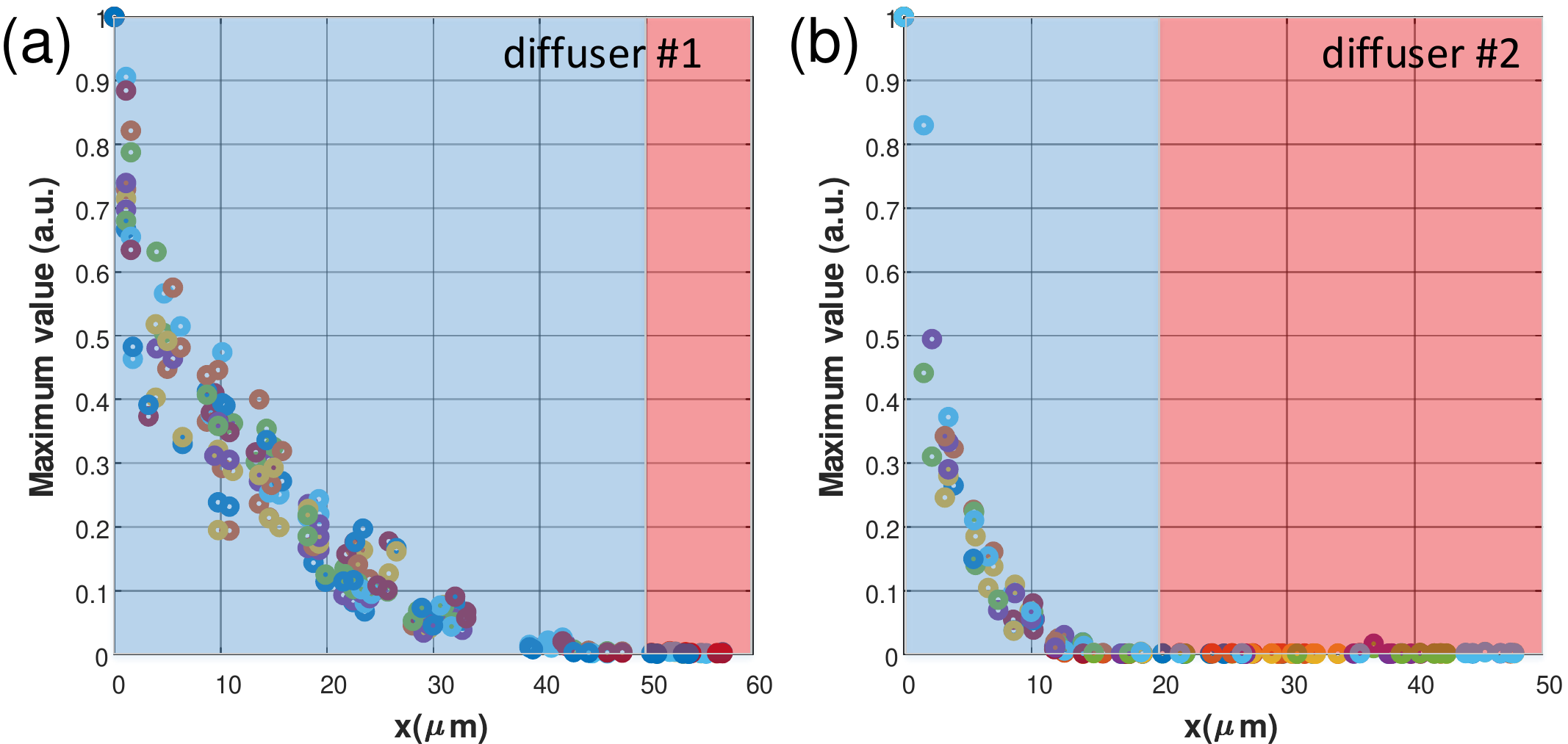}
\caption{\label{fig:8} (a)-(b) Plot of $\alpha= \frac {max\{o_{i,k}\}} {max\{o_{k,k}\}}$ as a function of relative distance between different fluorescent emitters for diffuser $\#1$ and diffuser $\#2$, where the colorcode represents the different data. }
\end{figure}

As described in the main text, it is possible to reconstruct the whole object as long as the ME patches have some overlaps. To present the details of FBR, we experimentally choose a fluorescent object which contains 3 fluorescent beads and display the various results from the pairwise deconvolution, $o_{i,k}$, and the different partial images, $O_{k}$. As shown in Fig.\ref{fig:7}, the $O_{k}$ of the emitter $k$ can be recovered by choosing $w_{k}$ as the PSF. By looking at the maximum value of those $o_{i,k}$, the shift $\vec{r}_{i,k}$ between fingerprints $w_{i}$ and $w_{k}$ can be retrieved, as shown in Fig.\ref{fig:1}c of the manuscript. In practice, the fingerprints coming from two emitters which are beyond the ME range will not provide the relative position information of their emitters (as they will be totally uncorrelated). Thus, it is necessary to infer whether the fingerprints $w_{i}$ and $w_{k}$ are within or beyond the ME range. In our method, we study $\alpha= \frac {max\{o_{i,k}\}} {max\{o_{k,k}\}}$ as a function of relative distance, where $max\{o_{i,k}\}$ stands for maximum value of $o_{i,k}$. A given threshold, $\alpha_{tres}$ of $\alpha$ is introduced to evaluate it. For example, if $\alpha$ is greater than $\alpha_{tres}$, $w_{i}$ and $w_{k}$ belong to the same ME range. Otherwise, they belong to different ME ranges, and their relative position cannot be retrieved. As shown in Fig.\ref{fig:8}, we experimentally investigate the maximum value of various results of pairwise deconvolutions as a function of distance between emitters. We set $\alpha_{tres}$ to $0.01$. The full workflow of technique is depicted in Alg.\ref{alg:a1}.

\begin{algorithm2e}[h!]
\DontPrintSemicolon
\SetAlgoLined
\KwInput{Series of camera images, $I_{fluo}(r,t)$.}
\KwOutput{Image of the object, $O^{Global}$.}
Estimate the rank ($\rho$) of the system from $I_{fluo}(r,t)$ (see Supplementary I).\;

Retrieve the spatial fingerprints ($w_{i}$) by using NMF with the estimated rank.\;

\For{$k=1,...,\rho$}{
Perform the pairwise deconvolution between $w_{k}$ and all the other fingerprints ($w_{i\neq k}$).\;

Retrieve the relative position between emitter $k$ and its neighbours inside the ME range.\;

Calculate the partial image of the object in the vicinity of the emitter ($O_{k}$) by adding the result of all the pairwise deconvolutions related to that emitter ($o_{i,k}$).\;
}

Merge all the partial images ($O_{k}$) into the final reconstruction ($O^{Global}$) using the relative position between emitters.\;

\caption{Image retrieval procedure}
\label{alg:a1}
\end{algorithm2e}

\section{Producing random illumination using a SLM}\label{Supplementary D}

\begin{figure}[h!]
\includegraphics[width=\linewidth]{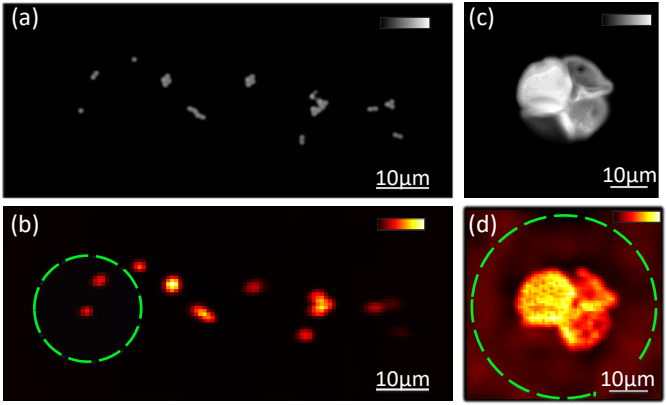}
\caption{\label{fig:9} Experimental results of imaging through scattering media by producing random illumination with a SLM. (a,c) are ground truths of fluorescent beads object and fluorescent continuous volumetric object. (c,d) are reconstructions corresponding to (a,b), respectively. The estimated rank $\rho$ of (c) and (d) is $14$ and $53$, respectively. $t = 10240$ for (c) and $t = 5120$ for (d) fluorescent speckle patterns are recorded. The dataset (c) is recorded with an exposure of $\SI{50}{\milli\second}$ and the exposure time of (d) is $\SI{20}{\milli\second}$. Dashed circle indicates the optical memory effect range.}
\end{figure}

\begin{figure}[h!]
\includegraphics[width=\linewidth]{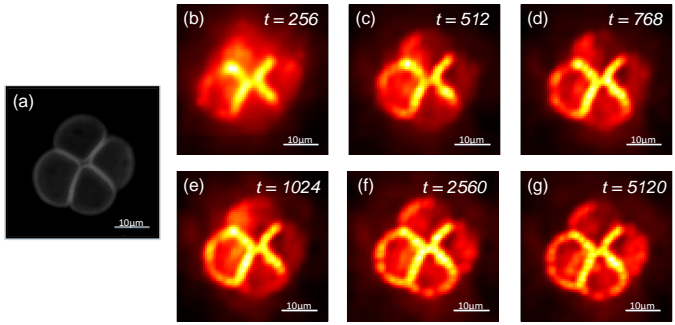}
\caption{\label{fig:10} Performance of reconstruction with different number of speckle patterns by producing random illumination with a SLM. (a) Ground truth of fluorescent continuous volumetric object taken without scattering medium. (c)-(d) Reconstructions with different number of illumination patterns. The rank of the NMF is estimated at $\rho = 27$ for (b), at $\rho = 31$ for (c), at $\rho = 33$ for (d), at $\rho = 36$ for (e),  at $\rho = 41$ for (f), and at $\rho = 53$ for (g). The dataset is recorded with an exposure of $\SI{20}{\milli\second}$.}
\end{figure}

Our technique retains the potential of employing a SLM for producing random illumination. The SLM can produce a large number of independent speckle illuminations. Experimentally, we replace the rotating diffuser with the SLM and perform experiments on fluorescent point-like objects and continuous volumetric objects by generating random illuminations with the SLM. The reconstruction is presented in Fig.\ref{fig:9}. To prove that it is possible to recover a more reliable image with more patterns, we show the reconstruction with the different number of patterns in Fig.\ref{fig:10}. Note that the results as shown in Fig.\ref{fig:9} and Fig.\ref{fig:10} are just tests to recover with the illumination patterns produced by SLM that allows generating a higher diversity of patterns than our current rotating diffuser, but it is not a limitation of our technique, as we could use different rotating diffusers to get more independent illumination patterns. 

\newpage
\section{Experimental setup}\label{Supplementary E}

\begin{figure}[h!]
\centering
\includegraphics[width=0.7\linewidth]{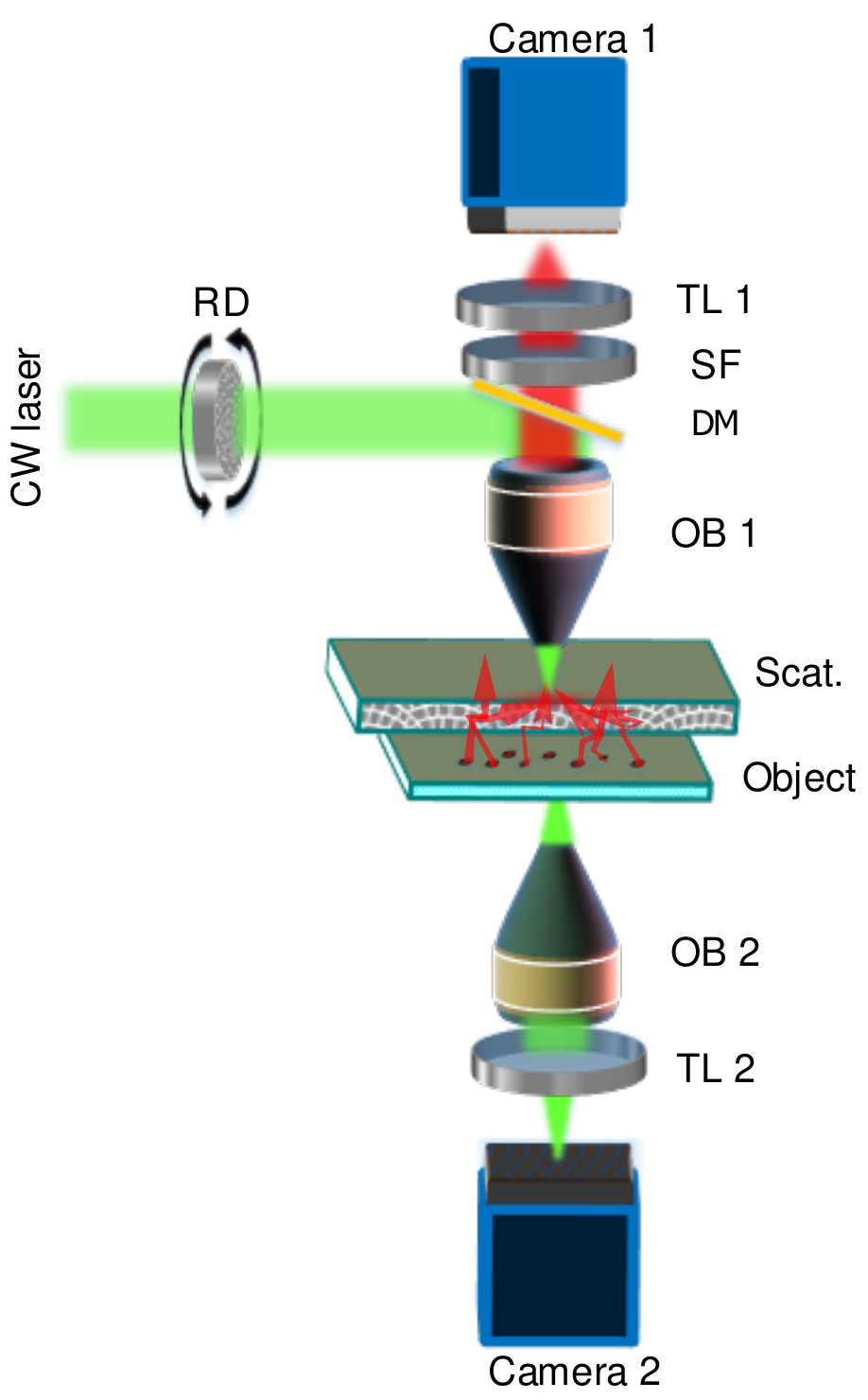}
\caption{\label{fig:11} Experimental setup. The expanded $\SI{532}{\nano\meter}$ laser beam illuminates the rotating diffuser and the modulated light is imaged on the back focal plane of objective 1 (OB1). The object is placed in the focal plane of OB1. Camera1 is located in the imaging plane of the fluorescent imaging system that is employed to capture fluorescent speckle. The passive controlling part is made of objective 2 (OB2), tube lens 2 (TL2), and camera 2. DM: dichroic mirror, SF: spectral filter, Scat.: scattering medium.}
\end{figure}

\clearpage
\bibliography{maintext}

\providecommand{\noopsort}[1]{}\providecommand{\singleletter}[1]{#1}%
\begin{thebibliography}{10}
\expandafter\ifx\csname url\endcsname\relax
  \def\url#1{\texttt{#1}}\fi
\expandafter\ifx\csname urlprefix\endcsname\relax\def\urlprefix{URL }\fi
\providecommand{\bibinfo}[2]{#2}
\providecommand{\eprint}[2][]{\url{#2}}

\bibitem{zhao_non-invasive_2001}
\bibinfo{author}{Zhao, M.}, \bibinfo{author}{Beauregard, D.~A.},
  \bibinfo{author}{Loizou, L.}, \bibinfo{author}{Davletov, B.} \&
  \bibinfo{author}{Brindle, K.~M.}
\newblock \bibinfo{title}{Non-invasive detection of apoptosis using magnetic
  resonance imaging and a targeted contrast agent}.
\newblock \emph{\bibinfo{journal}{Nature Medicine}}
  \textbf{\bibinfo{volume}{7}}, \bibinfo{pages}{1241--1244}
  (\bibinfo{year}{2001}).

\bibitem{artzi_vivo_2011}
\bibinfo{author}{Artzi, N.} \emph{et~al.}
\newblock \bibinfo{title}{In vivo and in vitro tracking of erosion in
  biodegradable materials using non-invasive fluorescence imaging}.
\newblock \emph{\bibinfo{journal}{Nature Materials}}
  \textbf{\bibinfo{volume}{10}}, \bibinfo{pages}{890--890}
  (\bibinfo{year}{2011}).

\bibitem{kozloff_non-invasive_2009}
\bibinfo{author}{Kozloff, K.~M.} \emph{et~al.}
\newblock \bibinfo{title}{Non-invasive optical detection of cathepsin
  {K}-mediated fluorescence reveals osteoclast activity in vitro and in vivo}.
\newblock \emph{\bibinfo{journal}{Bone}} \textbf{\bibinfo{volume}{44}},
  \bibinfo{pages}{190--198} (\bibinfo{year}{2009}).

\bibitem{Goodman1976}
\bibinfo{author}{Goodman, J.~W.}
\newblock \bibinfo{title}{{Some fundamental properties of speckle}}.
\newblock \emph{\bibinfo{journal}{Journal of the Optical Society of America}}
  \textbf{\bibinfo{volume}{66}}, \bibinfo{pages}{1145} (\bibinfo{year}{1976}).

\bibitem{Bender}
\bibinfo{author}{Bender, N.}, \bibinfo{author}{Yılmaz, H.},
  \bibinfo{author}{Bromberg, Y.} \& \bibinfo{author}{Cao, H.}
\newblock \bibinfo{title}{{Customizing speckle intensity statistics}}.
\newblock \emph{\bibinfo{journal}{Optica}} \textbf{\bibinfo{volume}{5}},
  \bibinfo{pages}{595} (\bibinfo{year}{2018}).

\bibitem{Abramson1978}
\bibinfo{author}{Abramson, N.}
\newblock \bibinfo{title}{{Light-in-flight recording by holography}}.
\newblock \emph{\bibinfo{journal}{Optics Letters}}
  \textbf{\bibinfo{volume}{3}}, \bibinfo{pages}{121} (\bibinfo{year}{1978}).

\bibitem{huang_optical_1991}
\bibinfo{author}{Huang, D.} \emph{et~al.}
\newblock \bibinfo{title}{Optical coherence tomography}.
\newblock \emph{\bibinfo{journal}{Science}} \textbf{\bibinfo{volume}{254}},
  \bibinfo{pages}{1178--1181} (\bibinfo{year}{1991}).

\bibitem{Mosk2012}
\bibinfo{author}{Mosk, A.~P.}, \bibinfo{author}{Lagendijk, A.},
  \bibinfo{author}{Lerosey, G.} \& \bibinfo{author}{Fink, M.}
\newblock \bibinfo{title}{{Controlling waves in space and time for imaging and
  focusing in complex media}}.
\newblock \emph{\bibinfo{journal}{Nature Photonics}}
  \textbf{\bibinfo{volume}{6}}, \bibinfo{pages}{283--292}
  (\bibinfo{year}{2012}).

\bibitem{rotter_light_2017}
\bibinfo{author}{Rotter, S.} \& \bibinfo{author}{Gigan, S.}
\newblock \bibinfo{title}{Light fields in complex media: {Mesoscopic}
  scattering meets wave control}.
\newblock \emph{\bibinfo{journal}{Reviews of Modern Physics}}
  \textbf{\bibinfo{volume}{89}}, \bibinfo{pages}{015005}
  (\bibinfo{year}{2017}).

\bibitem{Vellekoop2007}
\bibinfo{author}{Vellekoop, I.~M.} \& \bibinfo{author}{Mosk, A.~P.}
\newblock \bibinfo{title}{{Focusing coherent light through opaque strongly
  scattering media}}.
\newblock \emph{\bibinfo{journal}{Optics Letters}}
  \textbf{\bibinfo{volume}{32}}, \bibinfo{pages}{2309} (\bibinfo{year}{2007}).

\bibitem{Horstmeyer2015}
\bibinfo{author}{Horstmeyer, R.}, \bibinfo{author}{Ruan, H.} \&
  \bibinfo{author}{Yang, C.}
\newblock \bibinfo{title}{{Guidestar-assisted wavefront-shaping methods for
  focusing light into biological tissue}}.
\newblock \emph{\bibinfo{journal}{Nature Photonics}}
  \textbf{\bibinfo{volume}{9}}, \bibinfo{pages}{563--571}
  (\bibinfo{year}{2015}).

\bibitem{Katz2019}
\bibinfo{author}{Katz, O.}, \bibinfo{author}{Ramaz, F.},
  \bibinfo{author}{Gigan, S.} \& \bibinfo{author}{Fink, M.}
\newblock \bibinfo{title}{{Controlling light in complex media beyond the
  acoustic diffraction-limit using the acousto-optic transmission matrix}}.
\newblock \emph{\bibinfo{journal}{Nature Communications}}
  \textbf{\bibinfo{volume}{10}}, \bibinfo{pages}{1--10} (\bibinfo{year}{2019}).

\bibitem{Popoff2010}
\bibinfo{author}{Popoff, S.~M.} \emph{et~al.}
\newblock \bibinfo{title}{{Measuring the transmission matrix in optics: An
  approach to the study and control of light propagation in disordered media}}.
\newblock \emph{\bibinfo{journal}{Physical Review Letters}}
  \textbf{\bibinfo{volume}{104}}, \bibinfo{pages}{1--4} (\bibinfo{year}{2010}).

\bibitem{Hofer2019}
\bibinfo{author}{Hofer, M.} \& \bibinfo{author}{Brasselet, S.}
\newblock \bibinfo{title}{{Manipulating the transmission matrix of scattering
  media for nonlinear imaging beyond the memory effect}}.
\newblock \emph{\bibinfo{journal}{Optics Letters}}
  \textbf{\bibinfo{volume}{44}}, \bibinfo{pages}{2137} (\bibinfo{year}{2019}).

\bibitem{Freund1988}
\bibinfo{author}{Freund, I.}, \bibinfo{author}{Rosenbluh, M.} \&
  \bibinfo{author}{Feng, S.}
\newblock \bibinfo{title}{{Memory effects in propagation of optical waves
  through disordered media}}.
\newblock \emph{\bibinfo{journal}{Physical Review Letters}}
  \textbf{\bibinfo{volume}{61}}, \bibinfo{pages}{2328--2331}
  (\bibinfo{year}{1988}).

\bibitem{Yllmaz2019}
\bibinfo{author}{Yılmaz, H.} \emph{et~al.}
\newblock \bibinfo{title}{{Angular memory effect of transmission
  eigenchannels}}.
\newblock \emph{\bibinfo{journal}{Physical Review Letters}}
  \textbf{\bibinfo{volume}{123}}, \bibinfo{pages}{203901}
  (\bibinfo{year}{2019}).

\bibitem{Osnabrugge}
\bibinfo{author}{Osnabrugge, G.}, \bibinfo{author}{Horstmeyer, R.},
  \bibinfo{author}{Papadopoulos, I.~N.}, \bibinfo{author}{Judkewitz, B.} \&
  \bibinfo{author}{Vellekoop, I.~M.}
\newblock \bibinfo{title}{{Generalized optical memory effect}}.
\newblock \emph{\bibinfo{journal}{Optica}} \textbf{\bibinfo{volume}{4}},
  \bibinfo{pages}{886}.

\bibitem{Bertolotti2012}
\bibinfo{author}{Bertolotti, J.} \emph{et~al.}
\newblock \bibinfo{title}{{Non-invasive imaging through opaque scattering
  layers}}.
\newblock \emph{\bibinfo{journal}{Nature}} \textbf{\bibinfo{volume}{491}},
  \bibinfo{pages}{232--234} (\bibinfo{year}{2012}).

\bibitem{Katz2014}
\bibinfo{author}{Katz, O.}, \bibinfo{author}{Heidmann, P.},
  \bibinfo{author}{Fink, M.} \& \bibinfo{author}{Gigan, S.}
\newblock \bibinfo{title}{{Non-invasive single-shot imaging through scattering
  layers and around corners via speckle correlations}}.
\newblock \emph{\bibinfo{journal}{Nature Photonics}}
  \textbf{\bibinfo{volume}{8}}, \bibinfo{pages}{784--790}
  (\bibinfo{year}{2014}).

\bibitem{Ruan2020}
\bibinfo{author}{Ruan, H.}, \bibinfo{author}{Liu, Y.}, \bibinfo{author}{Xu,
  J.}, \bibinfo{author}{Huang, Y.} \& \bibinfo{author}{Yang, C.}
\newblock \bibinfo{title}{{Fluorescence imaging through dynamic scattering
  media with speckle-encoded ultrasound-modulated light correlation}}.
\newblock \emph{\bibinfo{journal}{Nature Photonics}}  (\bibinfo{year}{2020}).

\bibitem{Lichtman2005}
\bibinfo{author}{Lichtman, J.~W.} \& \bibinfo{author}{Conchello, J.~A.}
\newblock \bibinfo{title}{{Fluorescence microscopy}}.
\newblock \emph{\bibinfo{journal}{Nature Methods}}
  \textbf{\bibinfo{volume}{2}}, \bibinfo{pages}{910--919}
  (\bibinfo{year}{2005}).

\bibitem{mangeat_super-resolved_2021}
\bibinfo{author}{Mangeat, T.} \emph{et~al.}
\newblock \bibinfo{title}{Super-resolved live-cell imaging using random
  illumination microscopy}.
\newblock \emph{\bibinfo{journal}{Cell Reports Methods}}
  \textbf{\bibinfo{volume}{1}}, \bibinfo{pages}{100009} (\bibinfo{year}{2021}).

\bibitem{Hofer2018}
\bibinfo{author}{Hofer, M.}, \bibinfo{author}{Soeller, C.},
  \bibinfo{author}{Brasselet, S.} \& \bibinfo{author}{Bertolotti, J.}
\newblock \bibinfo{title}{{Wide field fluorescence epi-microscopy behind a
  scattering medium enabled by speckle correlations}}.
\newblock \emph{\bibinfo{journal}{Optics Express}}
  \textbf{\bibinfo{volume}{26}}, \bibinfo{pages}{9866} (\bibinfo{year}{2018}).

\bibitem{Boniface2020}
\bibinfo{author}{Boniface, A.}, \bibinfo{author}{Dong, J.} \&
  \bibinfo{author}{Gigan, S.}
\newblock \bibinfo{title}{{Non-invasive focusing and imaging in scattering
  media with a fluorescence-based transmission matrix}}.
\newblock \emph{\bibinfo{journal}{Nature Communications}}
  \textbf{\bibinfo{volume}{11}}, \bibinfo{pages}{6154} (\bibinfo{year}{2020}).

\bibitem{Berry2007}
\bibinfo{author}{Berry, M.~W.}, \bibinfo{author}{Browne, M.},
  \bibinfo{author}{Langville, A.~N.}, \bibinfo{author}{Pauca, V.~P.} \&
  \bibinfo{author}{Plemmons, R.~J.}
\newblock \bibinfo{title}{{Algorithms and applications for approximate
  nonnegative matrix factorization}}.
\newblock \emph{\bibinfo{journal}{Computational Statistics and Data Analysis}}
  \textbf{\bibinfo{volume}{52}}, \bibinfo{pages}{155--173}
  (\bibinfo{year}{2007}).

\bibitem{Moretti2020a}
\bibinfo{author}{Moretti, C.} \& \bibinfo{author}{Gigan, S.}
\newblock \bibinfo{title}{{Readout of fluorescence functional signals through
  highly scattering tissue}}.
\newblock \emph{\bibinfo{journal}{Nature Photonics}}
  \textbf{\bibinfo{volume}{14}}, \bibinfo{pages}{361--364}
  (\bibinfo{year}{2020}).

\bibitem{Pegard2016}
\bibinfo{author}{P{\'{e}}gard, N.~C.} \emph{et~al.}
\newblock \bibinfo{title}{{Compressive light-field microscopy for 3D neural
  activity recording}}.
\newblock \emph{\bibinfo{journal}{Optica}} \textbf{\bibinfo{volume}{3}},
  \bibinfo{pages}{517} (\bibinfo{year}{2016}).

\bibitem{Pnevmatikakis2016}
\bibinfo{author}{Pnevmatikakis, E.~A.} \emph{et~al.}
\newblock \bibinfo{title}{{Simultaneous Denoising, Deconvolution, and Demixing
  of Calcium Imaging Data}}.
\newblock \emph{\bibinfo{journal}{Neuron}} \textbf{\bibinfo{volume}{89}},
  \bibinfo{pages}{285} (\bibinfo{year}{2016}).

\bibitem{Antipa2018}
\bibinfo{author}{Antipa, N.} \emph{et~al.}
\newblock \bibinfo{title}{{DiffuserCam: lensless single-exposure 3D imaging}}.
\newblock \emph{\bibinfo{journal}{Optica}} \textbf{\bibinfo{volume}{5}},
  \bibinfo{pages}{1} (\bibinfo{year}{2018}).

\bibitem{hutchins_position-dependent_2008}
\bibinfo{author}{Hutchins, L.~N.}, \bibinfo{author}{Murphy, S.~M.},
  \bibinfo{author}{Singh, P.} \& \bibinfo{author}{Graber, J.~H.}
\newblock \bibinfo{title}{Position-dependent motif characterization using
  non-negative matrix factorization}.
\newblock \emph{\bibinfo{journal}{Bioinformatics}}
  \textbf{\bibinfo{volume}{24}}, \bibinfo{pages}{2684--2690}
  (\bibinfo{year}{2008}).

\bibitem{Daoud2017}
\bibinfo{author}{Wang, Z.}, \bibinfo{author}{Bovik, A.},
  \bibinfo{author}{Sheikh, H.} \& \bibinfo{author}{Simoncelli, E.}
\newblock \bibinfo{title}{{Image Quality Assessment: From Error Visibility to
  Structural Similarity}}.
\newblock \emph{\bibinfo{journal}{IEEE Transactions on Image Processing}}
  \textbf{\bibinfo{volume}{13}}, \bibinfo{pages}{600--612}
  (\bibinfo{year}{2004}).

\end{thebibliography}

\end{document}